\documentclass[12pt]{article}
\usepackage{amsmath}
\usepackage{amssymb}
\usepackage{cancel}
\usepackage{graphicx}
\usepackage{color}
\usepackage{wrapfig}

\renewcommand{\thefootnote}{\fnsymbol{footnote}}

\begin{document}
\baselineskip=19.5pt

\begin{titlepage}

\begin{flushright}
{\small MISC-2015-04}
\end{flushright}
\begin{center}
\vspace*{17mm}

{\large\bf%
Probing heavy neutrinos in the COMET experiment
}

\vspace*{10mm}
Takehiko~Asaka,$^{*}$
~~Atsushi~Watanabe$^{\dag}$ 
\vspace*{10mm}

$^*${\small {\it Department of Physics, 
Niigata University, Niigata 950-2181, Japan,}}\\

$^\dag${\small {\it Maskawa Institute for Science and Culture, Kyoto Sangyo University, 
Kyoto 603-8555, Japan
}}\\

\vspace*{3mm}

{\small (October, 2015)}
\end{center}

\vspace*{7mm}
\begin{abstract}\noindent%
We argue that the COMET experiment --- a dedicated experiment for the $\mu$-$e$ 
conversion search ---  can be a powerful facility to search for heavy neutrinos 
in the mass range $1\,{\rm MeV} \lesssim 
M \lesssim 100\,{\rm MeV}$.
The stopped muons captured by the target nuclei or decaying in orbit 
are efficiently produce heavy neutrinos via the active-sterile mixing.
The produced heavy neutrinos then decay to electron-positron pair (plus an
active neutrino), which events are clearly seen by the cylindrical drift chamber 
surrounding the target.
The expected sensitivity is comparable to the PS191 bound when the COMET 
experiment achieves $\sim 10^{17}$ stopping muons in the target.
\end{abstract} 

\end{titlepage}

\newpage
\renewcommand{\thefootnote}{\fnsymbol{footnote}}
\section{Introduction}
\label{intro}

Heavy neutrino is one of the most interesting new physics candidates. 
What makes this particle promising is the well-established
fact that the neutrinos are massive.
The most simple and natural way to account for the neutrino masses 
is introducing gauge-singlet fermions into the standard model.
In such a theory, the left-handed neutrinos $\nu_{L_\alpha}$ ($\alpha = e, \mu, \tau$)
are often mixed with the gauge-singlet fermions after the electroweak symmetry 
breaking, such that
\begin{eqnarray}
\nu_{L_\alpha} = \sum_{i=1}^{3}U_{\alpha i} \,\nu_i \, + \, \Theta_{\alpha} \nu_H.
\end{eqnarray}
Here $\nu_{L_\alpha}$ denote the flavor eigenstates of the left-handed neutrinos, 
$U_{\alpha i}$ is the Maki-Nakagawa-Sataka matrix, $\nu_i$ stand for the mass 
eigenstates of the ordinary neutrinos ($i = 1,2,3$),
$\nu_H$ is the heavy neutrino, and $\Theta_{\alpha} (|\Theta_\alpha| \ll 1)$ is 
the active-sterile mixing which rules the strength of the gauge interactions 
for $\nu_H$\footnote{The
extension to the multi-generation case is trivially done by replacing 
$\Theta_{\alpha} \nu_H$ with $\sum_j \Theta_{\alpha j} \nu_{H_j}$.
In this paper, we shall consider one heavy neutrino just for simplicity.}.

\vspace{2mm}

There is yet another motivation to consider the heavy neutrinos, namely,
the baryon asymmetry of the universe.
It is known that a (nearly) degenerate pair of heavy neutrinos in the mass range 
$1\,{\rm MeV} \lesssim M \lesssim 100\, {\rm GeV}$ can account for the baryon 
asymmetry of the universe through the oscillation taking place in the early 
universe~\cite{BAU,Asaka:2005pn,Shaposhnikov:2008pf,Asaka:2010kk,Canetti:2010aw,
Asaka:2011wq,Canetti:2012kh}.
Contrary to the standard leptogenesis~\cite{leptogenesis} with the heavy neutrino masses
around the grand unification scale, this mechanism predicts the heavy 
neutrinos testable by terrestrial experiments~\cite{Kusenko:2004qc,Gorbunov:2007ak,
H1,Asaka:2012hc,Asaka:2012bb,Aaij:2014aba,Dib:2014iga,
Boyanovsky:2014lqa,Blondel:2014bra,H2}. 

\vspace{2mm}

Having these strong motivations, SHiP~\cite{SHiP1,SHiP2} and DUNE~\cite{DUNE} are 
planning dedicated searches for heavy neutrinos.
These experiments will explore hitherto unexplored ranges of parameters far beyond 
the current bounds, as flagships of hidden particle searches in the coming decades.
Until physics run of these projects turn on, it would be desirable to have alternative 
searches with shorter-term ability. 
The heavy neutrinos are efficiently produced by muon and/or meson decay just like
the ordinary neutrinos. 
Thus a relevant question is if we can employ some existing or forthcoming
facilities in high intensity frontier.

\vspace{2mm}

In this paper, we focus on the COMET experiment~\cite{COMET} --- a dedicated experiment 
for the $\mu$-$e$ conversion search --- as an example of such an idea.
The COMET experiment plans to stop $\sim 10^{16}\,(10^{18})$ muons on the target 
in Phase-I~(Phase-II)~\cite{COMET}.
With these enormous numbers of muon, this experiment is potentially capable 
of discovering heavy neutrinos in unexplored parameter range beyond the 
strongest bound set by the PS191~\cite{PS191a,PS191b}.
The details are the followings.

\section{Expected sensitivity}
\label{events}
The COMET experiment searches the $\mu$-$e$ conversion process by looking for
the single electrons of $105\,{\rm MeV}$ energy coming out from the muonic atoms.
The experimental site is in the Hadron Experimental Hall of J-PARC.
The muon beam is produced from the pions following after the collision of
the bunched $8\,{\rm GeV}$ proton beam with a graphite target.
The pions are efficiently captured by the pion capture solenoid.
In Phase-I, the muons are stopped in an aluminium target after the
first $90^\circ$ bend of the muon transport solenoid.
The target is surrounded by the cylindrical drift chamber (CDC).
In Phase-II, the muon transport sector is extended and the stopping target 
is placed after the second $90^\circ$ bend.
The detector section is also extended so as to select the electron's momenta
by the electron transport solenoid.
See Ref.~\cite{COMET} for more details.
In what follows, we shall focus on the Phase-I detector setup.

\vspace{2mm}

About 60\% of the stopped muons are captured by the aluminium nuclei, changing 
themselves into the muon neutrinos $\nu_\mu$. 
The rest 40\% of the muons decay in orbit\footnote{The third branch of importance 
(in the normal discussion) is the $\mu$-$e$ conversion, but this is of course 
negligible in the study of the heavy neutrinos.}.
Suppose for simplicity that the heavy neutrino is in the mass range $1\,{\rm MeV} 
\lesssim M \lesssim 100\,{\rm MeV}$ and predominantly couples to muons, namely 
$|\Theta_\mu|^2\gg |\Theta_{e,\tau}|^2$.
In what follows, we focus on this parameter regime unless otherwise stated.
Then the daughter $\nu_\mu$ produced by the above two processes is ``replaced''
with the heavy neutrino $\nu_H$ at the rate $|\Theta_\mu|^2$.
Namely, when $N^{\rm stop}_\mu$ muons are stopped, 
$N^{\rm stop}_\mu  |\Theta_\mu|^2$ heavy neutrinos are produced\footnote{
In this paper, we do not consider the effect of the heavy neutrino mass in the 
production rate, since our  aim in this paper is estimating the ability of
the COMET experiment in comparison with the PS191 bounds.}. 
\begin{figure}[t]
\begin{center}
\scalebox{0.35}{
\includegraphics{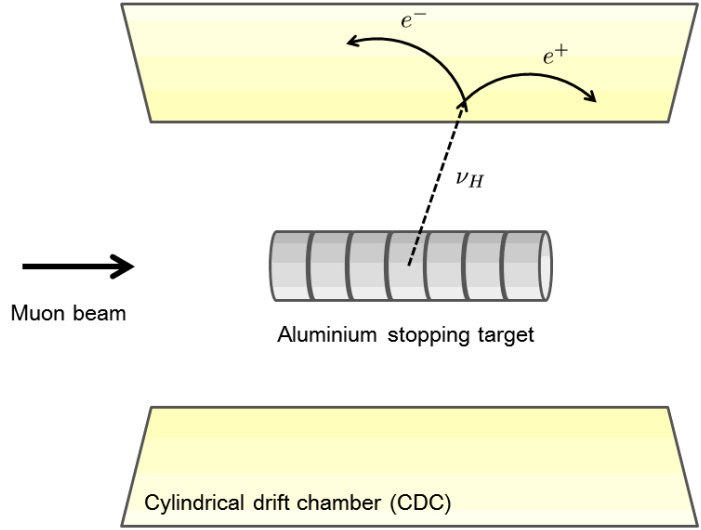}
}
\end{center}
\caption{A schematic view of the decay event of the heavy neutrino.}
\label{event}
\end{figure}
\vspace{2mm}

Within this parameter range, the main decay mode of $\nu_H$ is $\nu_H \to 3\nu$ 
and the subdominant mode is $\nu_H \to e^-e^+\nu$. 
The latter subdominant mode is detectable if the electron pair hit the CDC
with sufficient energies.
A schematic view of the decay event is shown in Fig.~\ref{event}.
Since the energy of the parent heavy neutrino $\nu_H$ is almost equal to
the muon mass $m_\mu = 105\,{\rm MeV}$, typical energy of each electron is
$\sim 35\,{\rm MeV}$.

\vspace{2mm}

\begin{table}[t]
\begin{center}
\begin{tabular}{lll}\hline\hline
Event selection & Value &  \\\hline
Geometrical acceptance & 0.7 &  \\
Timing window & 0.3/0.8& \\
Trigger efficiency& 0.8 & \\
Data acquisition system & 0.8 & \\
Track reconstruction & 0.8 & \\ \hline
Total & 0.11/0.29 &  \\ \hline\hline
\end{tabular}
\end{center}
\caption{Breakdown of the heavy neutrino signal acceptance.
For the timing window we examine two options of $0.3$ or $0.8$.}
\label{tab1}
\end{table}
The decay width of each process is given by
\begin{eqnarray}
&&\Gamma(\nu_H \to 3\nu) = \frac{G_F^2 M^5 |\Theta_\mu|^2}{192\pi^3},
\label{decay1}\\
&&\Gamma(\nu_H \to e^-e^+\nu) = \Gamma(\nu_H \to 3\nu)\left(
\frac{1}{4} - \sin^2\theta_W +2\sin^4\theta_W \right).
\label{decay2}
\end{eqnarray}
The fraction for the latter detectable mode is $1/4 - \sin^2\theta_W +2\sin^4
\theta_W = 0.13$
\footnote{Ref.~\cite{Asaka:2012hc}
presents more details on the kinematics for $e^\pm$ in the final state.
Those include the distributions of the invariant mass, the visible energy, 
the opening angle, etc.}.
Let us assume that the neutrinos are the Majorana particles.
Then the lifetime of $\nu_H$ is given by $\tau \simeq 
1/2\Gamma(\nu_H \to 3\nu)$ $=4.6 \times 10^{-5} \frac{1}{|\Theta_\mu |^{2}}
\left( \frac{50 \,{\rm MeV}}{M} 
\right)^5 \,{\rm (s)}$. 
The decay length for the detectable mode is $\lambda \simeq 1.1 \times 10^5 
\,\frac{\beta\gamma}{|\Theta_\mu |^{2}}\left( \frac{50 \,{\rm MeV}}{M} \right)^5  
\,{\rm (m)}$, where $\beta\gamma = \sqrt{m_\mu^2 - M^2}/M $.
When we pick up $M = 50\,{\rm MeV}$ and $|\Theta_\mu|^2 = 10^{-5}$ which are on 
the PS191 bound for example,
the signal decay length becomes $\lambda \simeq 2.0\times 10^{10}\,{\rm m}$.
The decay length is seemingly too long, and one may think the detection is a 
formidable task.
However, with a detector of $\mathcal{O}(1)\,{\rm m}$ length, the probability that
a heavy neutrino decay in the detector region is $1/\lambda \sim 10^{-10}$.
This means if we can gather $10^{11}$ heavy neutrinos $\mathcal{O}(10)$ events 
are expected to be observed.

\vspace{2mm}

\begin{figure}[t]
  \centerline{
  \includegraphics[scale=0.65]{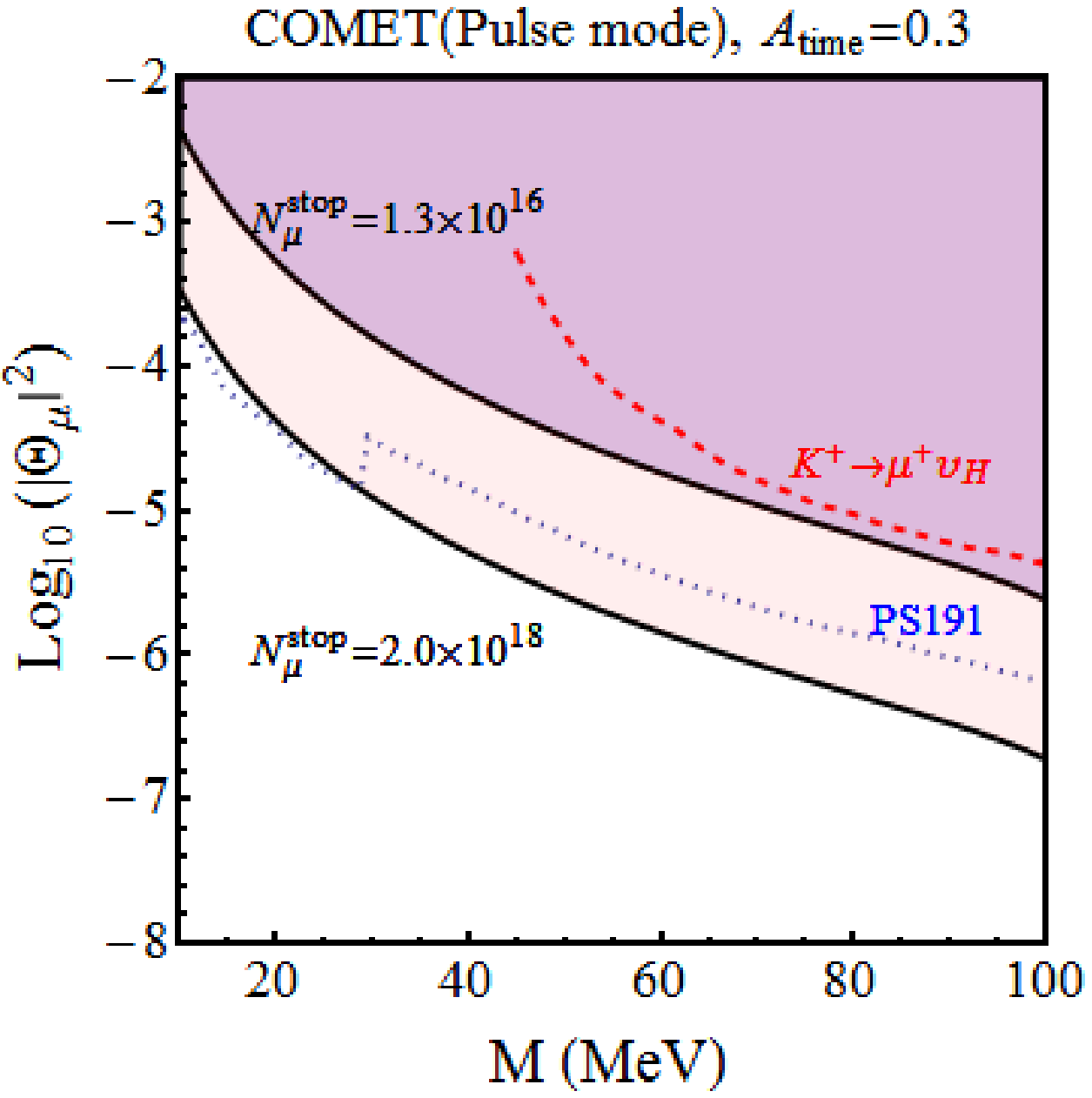}%
  \includegraphics[scale=0.65]{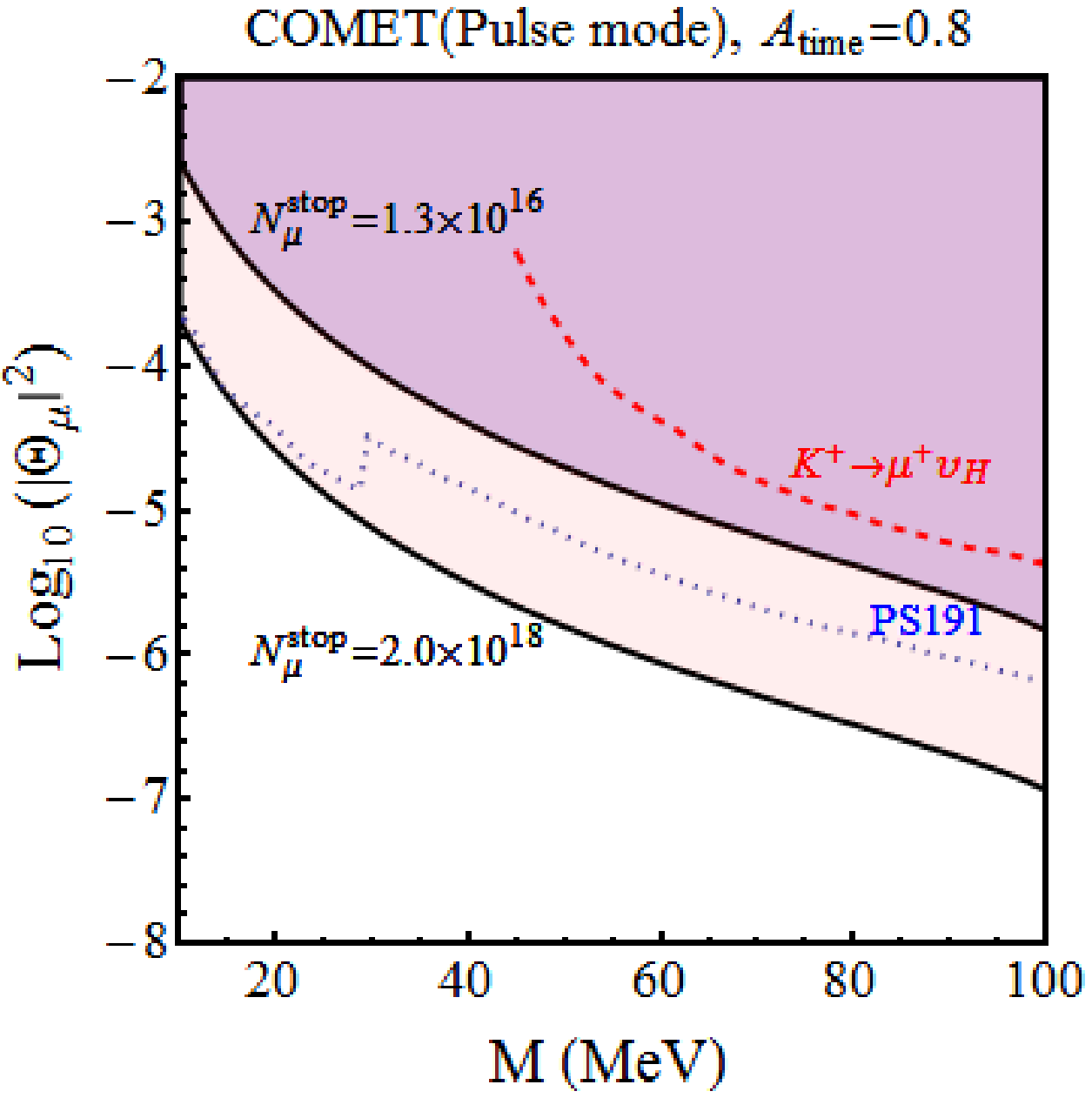}%
  }%
\caption{The 90\% CL limits when the COMET experiment observes null event.
The two solid curves show the COMET limits for $N^{\rm stop}_\mu = 1.3 \times
10^{16}$ and $N^{\rm stop}_\mu = 2.0 \times 10^{18}$.
The dotted curve labeled ``PS191'' is the 90\% CL limit placed by PS191~\cite{PS191a,
PS191b}.
The dashed curve labeled ``$K^+ \to \mu^+ \nu_H$'' is the bound placed by
the peak search in kaon decay~\cite{K2}.
The left panel shows the case where the timing acceptance $A_{\rm time}$ is
taken as $0.3$.
The right panel shows the case where $A_{\rm time} = 0.8$.}
\label{bounds}
\end{figure}

The number of events is estimated by
\begin{eqnarray}
{\rm Events}\,=\, N_{\mu}^{\rm stop} |\Theta_\mu|^2 
\frac{L}{\lambda} \,A,
\label{eventN}
\end{eqnarray}
where $N_{\mu}^{\rm stop}$ is the number of the muons stopped in the 
target, $L$ is the path length that heavy neutrinos across the CDC,
$\lambda$ is the decay length for the signal mode $\nu_H \to e^- e^+ \nu$,
and $A$ is the acceptance factor.
The path length $L$ depends on the production point and the flight direction
of the heavy neutrino.
In what follows we simply set $L = 0.3\,{\rm m}$, where $0.3\,{\rm m}$ is
the difference between the inner and outer radius of the CDC~\cite{COMET}.

\vspace{2mm}

The acceptance factor is the product of the several factors such as geometrical
acceptance, timing window, trigger efficiency, etc.
For these factors we use the same numbers as the $\mu$-$e$ conversion case~\cite{COMET}
except for the geometrical acceptance and the timing window.
For the geometrical acceptance, we assume that ``longitudinal heavy neutrinos'' 
will not leave detectable signals, where longitudinal heavy neutrinos mean 
those who are going through the up and down-stream ends that are not covered by the CDC.
This estimates 70\% of the heavy neutrinos are acceptable.
As for the timing window, we examine two options of $0.3$ and $0.8$, where $0.3$
is the same number as the $\mu$-$e$ conversion case. 
On the other hand, the number $0.8$ is meant as a potentially available number 
for the heavy neutrino search without the timing window, since it leaves 
characteristic $e^\pm$ pair which 
may not need a strict separation from the beam-related backgrounds.
The breakdown of the acceptance factor is presented in Tab.~\ref{tab1}\footnote{
A comment on the timing of the heavy neutrino signal:
The time of flight $\Delta t$ for $\nu_H$ is estimated by $\Delta t = r/c\beta$ 
where $r \simeq 0.5\,{\rm m}$ is the inner radius of the CDC. 
The time $\Delta t$ becomes significant only when the heavy neutrino
mass is in the vicinity of the muon mass. That is, $\Delta t > 100\,{\rm ns}$
for $\frac{m_\mu- M}{m_\mu} <1.4\times 10^{-4}$.
Thus, in most cases, the heavy neutrino events belong to the bursts
after $860\,{\rm ns}$ from the prompt timing~\cite{COMET}.}.

\vspace{2mm}

Fig.~\ref{bounds} shows the 90\% CL limits when the COMET experiment observes 
null event.
The left panel shows the case where the timing acceptance $A_{\rm time}$ is
taken as $0.3$, while the right panel shows the case where $A_{\rm time} = 0.8$.
The solid curves show the COMET limits in each panel. 
The dotted curve labeled ``PS191'' is the 90\% CL limit placed by PS191~\cite{PS191a,
PS191b}.
The dashed curve labeled ``$K^+ \to \mu^+ \nu_H$'' is the bound set by
the peak search in kaon decay~\cite{K2}.
Although the event estimation by Eq.~(\ref{eventN}) is applicable only for the Phase I
whose goal is $N^{\rm stop}_\mu = 1.3 \times 10^{16}$, we also put the curve 
for $N^{\rm stop}_\mu = 2.0 \times 10^{18}$ to get an idea how good the whole COMET 
project is, under the assumption that the Phase II setup can keep the same
performance for the heavy neutrino search.

\section{Discussions}
\label{discussions}
We conclude from Fig.~\ref{bounds} that the heavy neutrino search with the COMET 
experiment is an idea with good potential. 
It is important to note, however, that the curves in Fig.~\ref{bounds} 
are drawn under the assumption
that the heavy neutrinos exclusively contribute to the detections of $e^\pm$ pair.
A potential background is the $e^\pm$ pair creation by gamma rays.
According to Ref.~\cite{COMET}, the radiative muon capture
$\mu^- + A \to \nu_\mu + A' + \gamma$ and the radiative pion capture 
$\pi^- + A \to \gamma + A'$ are followed by $\gamma \to e^- + e^+$.
When these follow-up pair creations take place in the CDC volume, 
they mimic the heavy neutrino signal.
More detailed and precise analysis may thus need a thorough understanding of
the pair creation by gamma rays inside the CDC volume.

\vspace{2mm}

A possible way to reject these  background is selecting the directions of 
the $e^\pm$ momenta.
For the $e^\pm$ creation by gamma rays coming from the inner region than
the CDC volume, the momenta of $e^\pm$ tend to be outgoing for the momentum 
conservation.
On the other hand, the $e^\pm$ from the heavy neutrino decay can be emitted
to ingoing directions at significant rates. 
The typical $\gamma$ factor of $e^\pm$ in the $\nu_H$ rest frame is
$\langle \gamma_{e}\rangle = \frac{M/3}{m_e}$, whereas the gamma factor
of $\nu_H$ in the laboratory frame is $\gamma_{\nu_H} = \frac{m_\mu}{M}$.
Hence, roughly speaking, if $\langle \gamma_e \rangle > \gamma_{\nu_H}$,
namely if $M > \sqrt{3 m_e m_\mu} = 13\,{\rm MeV}$, then $e^\pm$ can head
in the opposite directions from the heavy neutrino momentum.
In such a mass regime distributions for the momentum direction may help to
reject the background.

\vspace{2mm}

If we relax the assumption $|\Theta_\mu|^2\gg |\Theta_{e,\tau}|^2$ so that 
the tau flavor mixing $\Theta_\tau$ takes part in the game,
the signal decay width becomes proportional to $\sqrt{|\Theta_\mu|^2 + |\Theta_\tau|^2}$
instead of $|\Theta_\mu|^2$.
In this case Fig.~\ref{bounds} should be read as a plot for the combination
$|\Theta_\mu|\sqrt{|\Theta_\mu|^2 + |\Theta_\tau|^2}$.
When we consider the full parameter space $\{\Theta_e, \Theta_\mu, \Theta_\tau, M\}$,
the thing gets more complicated, owing to the fact the signal decay $\nu_H 
\to e^- e^+ \nu$ is conducted by both of the charged and the neutral currents.
However, the electron component $|\Theta_e|$ is much more severely constrained
than the other two parameters~\cite{H1,H2}.
The plots in Fig.~\ref{bounds} (with the replacement mentioned above)
therefore cover all the cases of practical interests.

\subsection*{Acknowledgments}
The authors thank Prof.~Joe Sato for the useful discussions in 
start-up phase.
T.A. is supported by JSPS KAKENHI Grant Number 25400249, 26105508 and 15H01031.
A.W. is supported by JSPS KAKENHI Grant Number 25400249.
\bigskip



\end{document}